\documentclass[sigconf]{acmart}
\usepackage{balance}
\def\BibTeX{{\rm B\kern-.05em{\sc i\kern-.025em b}\kern-.08emT\kern-.1667em\lower.7ex\hbox{E}\kern-.125emX}}
\copyrightyear{2020}
\acmYear{2020}
\setcopyright{acmlicensed}
\acmConference[ASIA CCS '20]{Proceedings of the 15th ACM Asia Conference
on Computer and Communications Security}{October 5--9, 2020}{Taipei,
Taiwan}
\acmBooktitle{Proceedings of the 15th ACM Asia Conference on Computer and
Communications Security (ASIA CCS '20), October 5--9, 2020, Taipei, Taiwan}
\acmPrice{15.00}
\acmDOI{10.1145/3320269.3384760}
\acmISBN{978-1-4503-6750-9/20/10}

\usepackage{adjustbox}
\usepackage{balance}
\usepackage{float}
\usepackage{tikz}
\usepackage{pifont}
\usepackage{adjustbox}
\usepackage{tabularx}
\usepackage{multirow}
\newcommand*\circled[1]{\tikz[baseline=(char.base)]{\node[shape=circle,draw,inner
sep=1pt, thick] (char) {#1};}} 
\usepackage{shortcut}
\usepackage{color, colortbl}
\usepackage{listings}
\usepackage{subcaption}

\begin{document}
\fancyhead{}

\title{You shall not pass: Mitigating SQL Injection Attacks on Legacy Web Applications}

\author{Rasoul Jahanshahi}
\email{rasoulj@bu.edu}
\affiliation{%
	\institution{Boston University}
}

\author{Adam Doup\'e}
\email{doupe@asu.edu}
\affiliation{%
	\institution{Arizona State University}
}

\author{Manuel Egele}
\email{megele@bu.edu}
\affiliation{%
	\institution{Boston University}
}



\addauthnote{megele}{red!40}
\addauthnote{rasoul}{green!40}
\newcommand\vBuiltin{built-in PHP function}
\newcommand\vPhpExt{PHP extension}
\newcommand\vP{SQLBlock}
\newcommand\vwebapp{web app}
\newcommand\vWebapp{Web app}
\newcommand\vdbapi{database API}
\newcommand\vdbint{database interface}
\newcommand{\vcs}{call-stack}
\newcommand{\vdbal}{database access layer}
\newcommand{\vDbal}{Database access layer}
\newcommand{\vCore}{\circled{C}}
\newcommand{\vsone}{\circled{1}}
\newcommand{\vstwo}{\circled{2}}
\newcommand{\vsthree}{\circled{3}}
\newcommand{\vsfour}{\circled{4}}

\newcommand{\vSA}{\framebox{\texttt{SA}}}
\newcommand{\vPG}{\framebox{\texttt{PG}}}
\newcommand{\vPR}{\framebox{\texttt{PR}}}
\newcommand{\vPE}{\framebox{\texttt{PE}}}
\newcommand{\vDE}{\framebox{\texttt{DE}}}
\newcommand{\vcmsstat}{70.5\%}
\newcommand{\vwebstat}{38.4\%}
\newcommand{\vdataset}{11}
\newcommand{\vdockerstat}{5}

\newcommand{\vcG}{\cellcolor{green!10}}
\newcommand{\vcR}{\cellcolor{red!10}}

\newcommand{\vdsp}{22.60\%}
\newcommand{\vdup}{26.42\%}
\newcommand{\vdi}{9.96\%}

\newcommand{\voverhead}{4\%}

\newcommand{\cmark}{\ding{51}}
\newcommand{\xmark}{\ding{55}}

\begin{abstract}
	SQL injection (SQLi) attacks pose a significant threat to the security of web
applications.
Existing approaches do not support object-oriented programming that renders
these approaches unable to protect the real-world \vwebapp{}s such as
Wordpress, Joomla, or Drupal against SQLi attacks.
We propose a novel hybrid static-dynamic analysis for PHP web applications that limits
each PHP function for accessing the database.
Our tool, \vP{}, reduces the attack surface of the vulnerable PHP functions in a
web application to a set of query descriptors that demonstrate the benign
functionality of the PHP function.

We implement \vP{} as a plugin for MySQL and PHP. Our approach does not require
any modification to the \vwebapp{}.
We evaluate \vP{} on 11 SQLi vulnerabilities in Wordpress, Joomla, Drupal,
Magento, and their plugins. 
We demonstrate that \vP{} successfully prevents all 11 SQLi exploits
with negligible performance overhead (i.e., a maximum of 3\% on a
heavily-loaded web server).

\end{abstract}



\maketitle

\section{Introduction}\label{sec:intro}
The growing number of users for services such as social networks, news, online
stores, and financial services makes these services a tempting source of
sensitive information for attackers.
Symantec's recent report~\cite{symantecreport}, shows an increase of 56\% in
web attacks from 2017 to 2018. 
Moreover, according to Akamai~\cite{akamaireport}, 65.1\% of web attacks were
SQLi attacks.
SQLi is a type of code injection attack, where an attacker aims to execute
arbitrary SQL queries on a database. 
In 2018, the number of SQLi vulnerabilities discovered in the top four most
popular \vwebapp{}s (i.e., Wordpress, Joomla, Drupal, and Magento) increased
by 267\% compared to the prior year.
There has been a great deal of  research into identifying  SQLi vulnerabilities
and  defending against SQLi attacks on \vwebapp{}s. 
Proposed approaches used various techniques such as static
analysis~\cite{livshits2005finding,saferphp,  pixy2006, dahse2014usenix,
dahse2014}, dynamic analysis~\cite{bandhakavi2007candid, liu2009sqlprob, wasp,
medeiros2016hacking, medeiros2019septic, sun2008sqlprevent, buehrer2005using},
or a mix of static-dynamic analysis~\cite{halfond2005amnesia, naderi2015joza,
boyd2004sqlrand}. 
While static analysis approaches can be promising, static analysis cannot
determine whether input sanitization is performed correctly or
not~\cite{diglossia2013}.
If the sanitization function does not properly sanitize user-input, SQLi
attacks can still happen.
Moreover, to the best of our knowledge, prior static analysis approaches for
finding SQLi vulnerabilities in PHP \vwebapp{}s do not support Object-oriented
programming (OOP) code.
Such shortcomings in static analyses leave SQLi vulnerabilities undetected in
\vwebapp{}s such as Wordpress, Joomla, and Drupal that more than 40\% of active
websites use~\cite{w3cms}.
Prior dynamic analyses use taint analysis~\cite{wasp,huang2004} and comparison
of query parse trees ~\cite{su2006essence, bandhakavi2007candid,
buehrer2005using, sun2008sqlprevent, merlo2007automated, medeiros2019septic}
for detecting SQLi attacks on \vwebapp{}s.
Such dynamic analyses follow an incomplete definition of SQLi attacks where a SQLi
attack always alters the syntactic structure of an SQL query.
Ray et al.~\cite{ray2012defining} show that this incomplete definition in
CANDID~\cite{bandhakavi2007candid}, SQLCheck~\cite{su2006essence},
WASP~\cite{wasp}, and SQLPrevent~\cite{sun2008sqlprevent}  does not prevent
specific SQLi attacks and also blocks benign requests.
Other dynamic approaches have attempted to create a profile of the executed SQL
queries and enforce the profile at runtime~\cite{medeiros2019septic,
merlo2007automated}.
The profiles are a mapping between the parse tree of the benign issued SQL
queries and the PHP functions that issued the queries.
The profiles created by such approaches are too coarse-grained.
Particularly, modern and complex \vwebapp{}s such as Drupal and Joomla define
database APIs that perform all database operations.
Database APIs create SQL queries using the principle of the encapsulation that
allows the local functions to issue an SQL query to the database without passing
the SQL query as an argument.
In such cases, existing approaches map SQL queries to functions in the database
APIs instead of mapping to the function that uses database API for
communicating with the database. 
Hence, the prior approaches create a coarse-grained mapping that can allow an
attacker to perform mimicry SQLi attacks.
Specifically, Mereidos et al. in SEPTIC~\cite{medeiros2019septic} propose an
approach to block SQLi attacks inside the database. 
During training mode, SEPTIC records a profile that maps the parse trees  of
benign issued SQL queries to an identifier. 
SEPTIC generates the identifier in \texttt{mysql} and \texttt{mysqli}; two
database extensions in PHP for communicating with a MySQL database. 
The identifier is inferred from the PHP \vcs{} that issued a call to one of the
methods in the \texttt{mysql} or \texttt{mysqli} API for executing an SQL query
on the database (e.g., \texttt{mysql\_query}). 
The identifier is a sequence of functions in the PHP \vcs{} that pass the SQL
query as an argument.
In enforcement mode, SEPTIC checks the parse tree of the SQL query against the
profile it obtained during training mode. 
The design of SEPTIC leaves two unsolved challenges.  
(i) The strict comparison of the SQL query's parse tree against the profile
leads SEPTIC to reject a range of dynamic yet benign SQL queries, thus causing
false positives.
(ii) The coarse-grained mapping of SEPTIC's profile allows an attacker to
perform mimicry SQLi attacks successfully. 
The approach for creating the identifier in SEPTIC does not consider the fact
that \vwebapp{}s do not necessarily pass the SQL query as an argument.
As a result, SEPTIC assigns the SQL queries to a small set of functions in
database API as an identifier.
We evaluated SEPTIC's protection model against the Drupalgeddon vulnerability
in Drupal~\cite{drupalgeddon}.
Database API in drupal uses the encapsulation concept that means Drupal's
functions do not pass the SQL query as an argument to the database API. 
Hence, SEPTIC maps all issued SQL queries to the same sequence of functions in
the database API instead of the function that interacts with the database
through the database API.
During training mode, SEPTIC creates its profile by mapping all the received
SQL queries to a single identifier.
This mapping in the profile means that any function that communicates with the
database in Drupal can issue all the SQL queries in the SEPTIC's profile.
For instance, an attacker can exploit the Drupalgeddon vulnerability in the
presence of the SEPTIC and use the login functionality to issue an SQL query for
creating an admin user.
Considering the challenges and open problems with existing defenses
against SQLi attacks for PHP \vwebapp{}s, we propose a novel hybrid
static-dynamic analysis and its implementation \vP{} to defend OOP \vwebapp{}s
against SQLi attacks. 
\vP{} consists of four steps for defending \vwebapp{}s against SQLi attacks.
In the first step, \vP{} collects benign inputs through unit tests or benign
browsing of \vwebapp{}s and creates a mapping between the issued SQL query and
the function that issued the query. 
The static analysis is necessary to determine the database API precisely and
subsequently identify the PHP function that uses this API to communicate with
the database correctly.
In the next step, \vP{} creates a profile based on the issued query from each
function in the \vwebapp{} during the training mode.  
The profile in \vP{} is a mapping between the function that issues the SQL
query and a query descriptor that describes the benign functionality of the SQL
query.
In the last step, \vP{} enforces the profile inside the database to prevent the
execution of any SQL query that does not match the profile at the runtime.
We evaluate our system on a total of \vdataset{} known SQLi vulnerabilities of
the top four most popular real-world \vwebapp{}s Wordpress, Drupal, Joomla,
Magento, and their plugins.  
\vP{} defends against all SQLi exploits, while SEPTIC can only defend against
four SQLi attacks in our dataset.
In summary, we make the following contributions: 

\begin{itemize}	
	\item 
		We recognize that the object-oriented programming paradigm
		poses challenges for existing systems that lead to false
		positives and reduced protection against SQLi attacks.
		We propose a novel system to statically and precisely identify
		database API in a PHP \vwebapp{}, and dynamically restrict the
		SQL queries that MySQL executes based on the PHP function that
		composes the SQL query.
	\item We present a prototype implementation called \vP{} as a MySQL
		plugin. It can be used with minimal modifications to MySQL
		for defending against more types of SQLi attacks against PHP
		\vwebapp{}s than prior work. (more details in \S~\ref{sec:relatedwork})
	\item We evaluate \vP{} for its security and performance
		characteristics on four popular PHP \vwebapp{}s and seven plugins.
		\vP{} protects the database and the \vwebapp{} against
		\vdataset{} previously known SQLi vulnerabilities in our
		evaluation dataset with an acceptable performance overhead (<3\%).
\end{itemize}
We will open source our implementation of \vP{}, including the testing and
evaluation dataset. Our dataset includes \vdataset{} vulnerable PHP
\vwebapp{} and plugins, as well as automated Selenium scripts recorded from human
interactions with each \vwebapp{}.

\section{Background}\label{sec:background}

In this section, we provide an overview of object-oriented programming in
PHP and the PHP extensions that are used to communicate with MySQL databases.
Afterward, we discuss MySQL and its plugin architecture.
 Understanding the OOP model in PHP is necessary for our static analysis. 
Besides that, the knowledge of database extensions in PHP for communicating
with MySQL and the power of MySQL plugins shapes the implementation of \vP{}. 
We then discuss different types of SQLi attacks that impact the profile created
in step \vsthree{} of \vP{}.

\subsection{PHP} \label{subsec:back:php}
PHP is an open-source server-side scripting language.
According to W3Techs~\cite{w3marketshare}, 79.1\% of all websites use PHP as
their server-side language. 
PHP supports binary extensions called \emph{plugins} that provide PHP with
additional features such as cryptographic algorithms, mail transfer, or
database communications. 
Database API in PHP provides an interface for communicating with a database.
Database API can be database-specific such as MySQL and SQLite, or a general
interface such as PHP Data Objects (PDO) for accessing various databases.
The \textit{mysqli} extension provides functionality to access MySQL databases
in PHP scripts.
Compared to \textit{mysql}, which is another PHP extension for accessing MySQL,
\textit{mysqli} provides three additional features: support for prepared
statements, multiple statement queries, and transactions.
PHP \vwebapp{}s tend to use \texttt{mysqli} due to aforementioned additional
capabilities. 

PDO is an abstraction layer that provides a consistent API for accessing
databases regardless of the database type. 
This feature allows a PHP script to use the same piece of PHP code to connect to
different types of databases and issue queries. 
Although PDO delivers a clean and simple API for accessing the database, it
only provides generic query-building functionality. 
For instance, PDO neither supports multiple SQL queries in one string,
asynchronous queries, nor automatic cleanup with persistent connections.
PHP supports the Object-Oriented Programming model, which introduces three new
concepts for developing PHP \vwebapp{}s: inheritance, polymorphism, and
encapsulation. 
Inheritance and polymorphism let developers extend the functionality of classes
or implement an interface in more than one way. 
Encapsulation bundles data and methods into a single unit. 
Hence, OOP allows developers to create modular programs and extend the
functionality of PHP database extension.
Additionally, PHP provides dynamic features, such as creating objects from
dynamic strings.
\texttt{new} is the keyword for creating objects from a class in PHP.
The argument for the \texttt{new} keyword, can be a class name or a string that
represents the name of the class.
An example is shown in Figure~\ref{fig:codesample}, line~\ref{line:instnew},
where the value of \texttt{getDriver()} defines the class that should be
instantiated.
Besides the object-oriented design of database APIs, PHP \vwebapp{}s  also
implement database procedures.
Database procedures handle instantiating objects from the database API and
return an object from the database API or a sub-type of the database API.
Throughout this paper, we call the \vdbapi{} and procedures as the \vdbal{}.
The \vdbal{} in the \vwebapp{} handles the communication of the \vwebapp{}'s
modules with the database. 
\vP{} determines the \vdbal{} in PHP \vwebapp{}s by reasoning about the source
code of the PHP \vwebapp{} statically with respect to the OOP implementation of
the \vwebapp{}s.

\subsection{MySQL} 
MySQL is an open-source database management system.
As of August 2019, according to Datanyze~\cite{mysqlmarketshare}, MySQL is used
in 46.03\% of the deployed websites on the Internet. 
MySQL supports a plugin API that enables developers to extend the functionality
of MySQL.
MySQL Plugins can implement user authentication, query rewriting components, or
new parsers for additional keywords and capabilities.
MySQL plugins have access to different data structures, depending on their
role. 
Of particular interest to this paper is the query rewrite plugin, which can
examine and modify a query when MySQL receives the query before execution. 
Query rewrite plugin has access to the parse tree of the SQL query that MySQL
received.
Each node in the parse tree based on its type contains information regarding
the element it represents from the SQL query.
For instance, the function node (e.g., \texttt{IN}, \texttt{<}) contains
information regarding the number of arguments passed to the SQL function.
\vP{} uses the information that each node contains during its training and
enforcement.
Postparse plugins also have access to the information regarding the type of the
SQL query (e.g., SELECT, INSERT) and the name of the table that the SQL query
needs to access. 
\vP{} uses the information above to create and enforce the query descriptors
for each received SQL query.

\subsection{SQL Injection attacks}
SQL injection (SQLi) is a code injection attack in which an attacker is able to
control a SQL query to execute malicious SQL statements to manipulate the
database.
SQLi attacks are classified into eight
categories~\cite{halfond2006classification, dahse2014usenix}:

\begin{enumerate} 
	
	\item \textit{\textbf{Tautologies:}} 
		The attacker injects a piece of code into the conditional
		clause (i.e., where clause) in a SQL query such that the
		SQL query always evaluates to
		true~\cite{halfond2006classification}.  
		The goal of this attack varies from bypassing authentication to
		extracting data depending on how the returned data is used in
		the application.
	
	\item \textit{\textbf{Illegal/Logically incorrect Queries:}} 
		By leveraging this vulnerability, an attacker can modify the
		SQL query to cause syntax, type conversion, or logical
		errors~\cite{halfond2006classification}. 
		If the \vwebapp{}'s error page shows the database error,
		the attacker can learn information about the back-end
		database.
		This vulnerability can be a stepping stone for further attacks
		that reveals the injectable parameters to the attacker.   

	\item \textit{\textbf{Union Query: }} 
		In union query attacks, the attacker tricks the application to
		append data from the tables in the database for a given
		query~\cite{halfond2006classification}. 
		An attacker adds one or more additional \textit{SELECT} clause,
		which start with the keyword \textit{UNION}, that leads to
		merging results from other tables in the database to the result
		of the original SQL query. 
		The goal of such an attack is to extract data from additional
		tables in the database.

	\item \textit{\textbf{Piggy-backed Query:}} 
		Piggy-backed query enables attackers to append at least one
		additional query to the original query.
		Therefore the database receives multiple queries in one string
		for execution~\cite{halfond2006classification}. 
		The attacker does not intend to modify the original query but
		to add additional queries.
		Using the piggy-backed query, an attacker can insert, extract,
		or modify data as well as execute remote commands as well as
		extract data from the database. 
		The success of the attack depends on if the database allows the
		execution of multiple queries from a single string.

	\item \textit{\textbf{Stored procedures:}} 
		Stored procedures are a group of SQL queries that encapsulate
		a repetitive task.
		Stored procedures also allow interaction with the operating
		system~\cite{halfond2006classification}, which can be invoked
		by another application, command line, or another stored
		procedure. 
		While a database has a set of default stored procedures, the
		SQL queries in a stored procedure can be vulnerable similar to
		SQL queries outside the stored procedure.

	\item \textit{\textbf{Inference:}} 
		In this type of attack, the application and the database are
		prevented from returning feedback and error messages;
		therefore, the attacker cannot verify whether the injection was
		successful or not~\cite{halfond2006classification}. 
		In the inference attacks, the attacker tries to extract data based
		on answers to true/false questions about the data already stored
		in the database.

	\item\textit{\textbf{Alternate Encoding:}} 
		In order to evade detection, the attackers use different encoding
		methods to send their payload to the database.  Each layer of
		the application deploys various approaches for handling
		encodings~\cite{halfond2006classification}. 
		The difference between handling escape characters can help an
		attacker to evade the application layer and execute an
		alternate encoded string on the database layer.

	\item \textit{\textbf{Second order injections:}} 
		One common misconception is that the data already stored in the
		database is safe to extract~\cite{dahse2014usenix}. 
		In a second order attack, an attacker sends his crafted SQL
		query to the database to store his attack payload in the
		database. 
		The malicious payload stays dormant in the database until the
		database returns it as a result of another query, and the
		malicious payload is insecurely used to create another SQL
		query. 
\end{enumerate}

\section{Related Work}\label{sec:relatedwork}

In this section, we review the relevant literature on defending \vwebapp{}s
against SQLi attacks.
We also compare \vP{} with five existing approaches and explain why prior
systems are not sufficient for PHP \vwebapp{}s that utilize OOP to communicate
with databases.

\newcommand\vdefhalf{%
	\begin{tikzpicture}		            
	\fill[black] (0,0) circle (0.75ex);
	\clip (-1ex,-1ex) rectangle (0,0.75ex);
	\fill[white] (0,0) circle (0.75ex);
	\end{tikzpicture}}
\newcommand\vdefy{\tikz\fill[fill=black] (0,0) circle (0.75ex);}
\newcommand\vdefn{\tikz\filldraw[color=black,fill=white] (0,0) circle (0.75ex);}

\newcommand\vAtool{\vP{}}
\newcommand\vBtool{SEPTIC~\cite{medeiros2019septic}}
\newcommand\vCtool{SQLrand~\cite{boyd2004sqlrand}}
\newcommand\vFtool{DIGLOSSIA~\cite{diglossia2013}}
\newcommand\vGtool{SQLCheck~\cite{su2006essence}}
\newcommand\vHtool{Merlo et. al.~\cite{merlo2007automated}}

\newcommand\vAatt{Taut.}
\newcommand\vBatt{Illegal/Incorrect}
\newcommand\vCatt{Union}
\newcommand\vDatt{Piggy-back}
\newcommand\vEatt{Stored proc.}
\newcommand\vFatt{Infer.}
\newcommand\vGatt{Alt. encoding}
\newcommand\vHatt{Second order inj.}

\begin{table}[b]
\resizebox{\columnwidth}{!}{
	\begin{tabular}{l|c|c|c|c|c|c|c|c|c}

		Tool & \rotatebox{90}{\vAatt{}} & \rotatebox{90}{\vBatt{}} & \rotatebox{90}{\vCatt{}} & \rotatebox{90}{\vDatt{}} & \rotatebox{90}{\vEatt{}} & \rotatebox{90}{\vFatt{}} & \rotatebox{90}{\vGatt{}} & \rotatebox{90}{\vHatt{}} & \rotatebox{90}{flagged as attack}\\ \hline
		
		\vCtool{} & \vdefy{} & \vdefn{} & \vdefy{} & \vdefy{} & \vdefn{} &\vdefn{} & \vdefn{} & \vdefn{} & 0 \\
		\vGtool{} &  \vdefhalf{}&\vdefhalf{}&\vdefhalf{}&\vdefhalf{}&\vdefn{}&\vdefhalf{}&\vdefhalf{} & \vdefn{} & 0\\ 
		\vHtool{} &  \vdefhalf{}& \vdefn&\vdefhalf{}&\vdefhalf{}&\vdefhalf{}&\vdefhalf{}&\vdefhalf{}&\vdefn{} & 0\\
		\vBtool{} &  \vdefhalf{}& \vdefn&\vdefhalf{}&\vdefhalf{}&\vdefhalf{}&\vdefhalf{}&\vdefhalf{}&\vdefhalf{} & 4\\	
		\vFtool{} &  \vdefy{}&\vdefy{}&\vdefy{}&\vdefy{}&\vdefn{}&\vdefy{}&\vdefn{}&\vdefn{} & 5\\
		\vAtool{} &  \vdefy{}&\vdefn{}&\vdefy{}&\vdefy{}&\vdefy{}&\vdefy{}&\vdefy{}&\vdefy{} & 11\\
	\hline
	\end{tabular}
	}
	\caption{Comparison of \vP{} with other techniques with respect to SQLi attack type. \vP{} provides the most effective protection.}
	\label{tab:compare}
	
\end{table}

Our comparison based on the SQLi attack type is presented in
Table~\ref{tab:compare}. 
For each SQLi attack type in Table~\ref{tab:compare}, \vdefy{} means the tool
can defend against the type of attack, \vdefn{} means the tool is ineffective,
and \vdefhalf{} means that the tool can partially defend the \vwebapp{} against
SQLi attack.
Partially defending means that either the tool can only defend \vwebapp{}s that
do not use OOP for implementing the communication with the database, or the
definition of SQLi attacks in the tool is incomplete.
The last column of Table~\ref{tab:compare} shows the number of SQLi exploits
from our dataset in Table~\ref{tab:defense} that each tool can prevent. 
\textit{Static Analysis}: Several proposed approaches focus on detecting
injection vulnerabilities statically in the source code of web
applications~\cite{dahse2014,pixy2006,huang2004,dahse2014usenix,sound2007}.
Dahse et al.~\cite{dahse2014} proposed RIPS, an inter- and intra-procedural
data flow analysis for detecting XSS and SQLi vulnerabilities in \vwebapp{}s.  
Pixy~\cite{pixy2006} implements a flow-sensitive data flow analysis to find XSS
and SQLi vulnerabilities in \vwebapp{}s. 
WebSSARI~\cite{huang2004} uses taint analysis to track untrusted user-inputs to
detect command injection vulnerabilities. 
Dahse et al.~\cite{dahse2014usenix} implement a context-sensitive taint
analysis to analyze read and write operations to the memory locations in
webserver for finding the second-order injections.
Wassermann et al.~\cite{sound2007} proposed a static analysis for detecting the
injection vulnerabilities in \vwebapp{}s.
A major drawback of prior analyzes are the inability to detect SQLi
vulnerabilities in \vwebapp{}s such as Wordpress, Joomla, and Drupal that
utilize OOP for communicating with databases.

\setcounter{topnumber}{4}
\begin{figure*}[h!]
	\includegraphics[width=\textwidth]{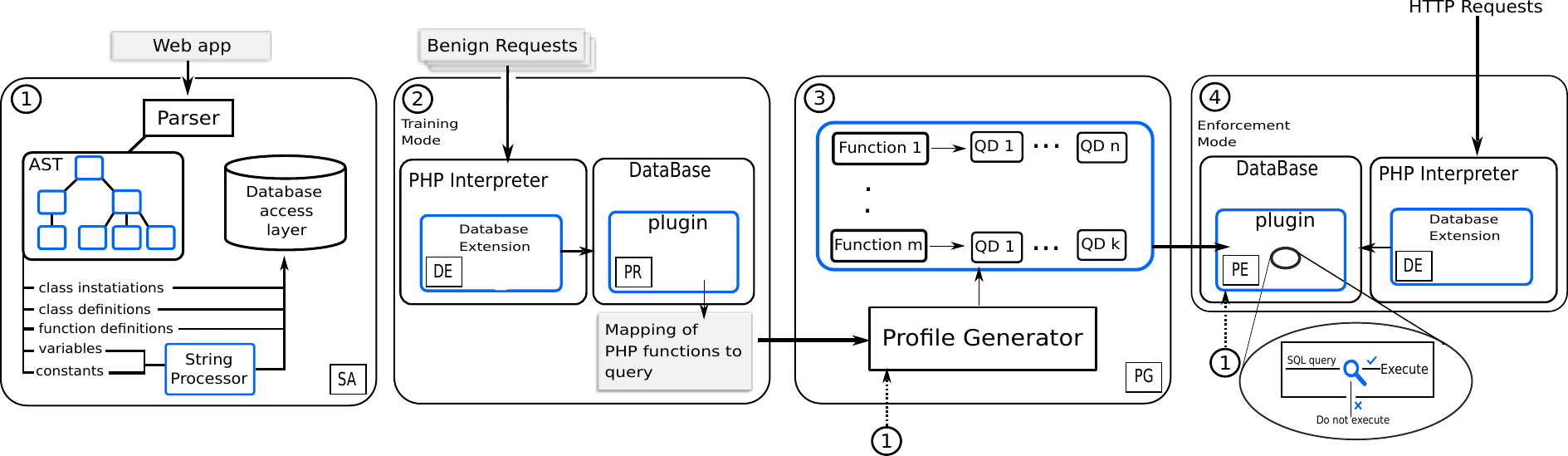}
	\caption{\vP{} extracts the database access layer of the \vwebapp{}, builds a
	mapping between the function and the SQL queries it issues, creates a
	profile for each function in the \vwebapp{} and enforces the profile
	using a MySQL plugin.}
\label{fig:sys}
\end{figure*}
\textit{Dynamic Analysis}:
Dynamic approaches track user-inputs~\cite{sun2008sqlprevent, buehrer2005using,
su2006essence, bandhakavi2007candid}, or build a profile of benign SQL
queries~\cite{medeiros2019septic, diglossia2013, medeiros2016hacking,
merlo2007automated} to prevent  SQLi attacks on \vwebapp{}s.
SQLPrevent~\cite{sun2008sqlprevent} analyzes generated queries for the
existence of HTTP request parameters and raises an alert when an HTTP request
parameter modifies the syntax structure of a query.
SQLGuard~\cite{buehrer2005using} proposed a dynamic approach for comparing the
parse tree of issued queries at runtime before and after the inclusion of user
inputs. SQLGuard needs to modify the source code of the \vwebapp{}. 
WASP~\cite{wasp} proposes a taint analysis to detect SQLi attacks on
\vwebapp{}s.
CANDID~\cite{bandhakavi2007candid} records a set of benign SQL queries that the
\vwebapp{} can issue by instrumenting the \vwebapp{}'s source code and
dynamically executing the SQL statements with benign inputs.
CANDID, SQLGaurd, WASP, and SQLPrevent assume that if the input does not change
the syntax structure of a SQL query, then a SQLi attack has not occurred.
Such an assumption can leave the \vwebapp{} vulnerable to SQLi attacks and also
blocks benign generated queries~\cite{ray2012defining}.
Unlike  CANDID, SQLGaurd, WASP, and SQLPrevent, \vP{} does not detect SQLi
attacks based on the modification to the syntax structure of the SQL query.
\vP{} generates a set of query descriptors for benign queries that each PHP
function issues to the database.
\vP{} allows functions in the \vwebapp{} to issue queries, as long as the query
matches its query descriptors.
Beyond this, \vP{} does not need to modify the source code of the \vwebapp{}
for its operation.
SQLCheck~\cite{su2006essence} tracks user-inputs to SQL queries and flags a SQL
query as an attack if user-input modifies the syntactic structure of the SQL
query.
This incomplete definition of SQLi attacks prevents SQLCheck from defending
against tautology, inference, stored-procedure, and alternate encoding attacks.
These four attacks do not necessarily modify the syntax structure
of a SQL query.
Considering this weaknesses, SQLCheck cannot protect \vwebapp{}s against any of
the vulnerabilities in our dataset mentioned in Table~\ref{tab:defense}.
Diglossia~\cite{diglossia2013} proposed a dual parser as an extension to the
PHP interpreter.
Diglossia maps the query without user-inputs to a shadow query, and then it
checks whether the parse tree of actual query and the shadow query are
isomorphic or not.
If both parse trees are isomorphic and the code in the shadow query is not
tainted with user-inputs, Diglossia passes the query to the back-end database.
Diglossia is unable to defend against Second-order injection since Diglossia
only checks queries with user-inputs.
Moreover, Diglossia cannot detect alternate-encoding and stored-procedure
attacks since these attacks do not modify the parse tree of the SQL
query~\cite{medeiros2019septic}. 
As shown in Table~\ref{tab:compare}, \vP{} defends \vwebapp{}s against more
variants of SQLi attacks than Diglossia.
SEPTIC~\cite{medeiros2019septic} creates a profile for each issued query during
the training phase and enforces this profile to protect \vwebapp{}s against
SQLi attacks. 
During the training, SEPTIC creates a query model that includes all the nodes
in the parse tree of a SQL query. 
The profile in SEPTIC is a mapping between the query model and an ID.
The ID is the sequence of functions that pass the query as an argument.
During the enforcement, SEPTIC uses this sequence of functions as an identifier
and finds the appropriate query model in the profile.
If the issued query matches the query model in the profile, SEPTIC allows the
database to execute the query.
Enforcing a profile based on the exact model of the generated queries that
includes the name of table columns and number of SQL functions prevents
\vwebapp{}s from generating dynamic yet benign SQL queries, which causes false
positives in SEPTIC.
For instance, assume there is a webpage for searching for published music
albums and users can search based on the name of an album, an artist's name, or
the released year.
If SEPTIC is trained with SQL queries that only includes the album's name or
the released year, it rejects any SQL queries from a user that searches using
the artist's name.
\vP{} solves this problem by creating query descriptors for SQL queries.
Query descriptors generalize the benign SQL queries, which allows the
\vwebapp{} to produce a range of dynamic queries.
Furthermore, to create an identifier for each issued query in the profile,
SEPTIC uses the information in the PHP \vcs{} that issued the call to methods
from \textit{mysql} or \textit{mysqli}.
SEPTIC checks the sequence of functions in the PHP \vcs{} for the presence of
SQL query in function's arguments.
Since OOP \vwebapp{}s do not pass the SQL query as an argument, SEPTIC cannot
generate a correct identifier for SQL queries.
Instead, it creates the same identifier for all the issued queries in the OOP
\vwebapp{}.
Consequently, an attacker can use a vulnerable function in the \vwebapp{} to
issue any query from the profile.
Considering the coarse-grained mapping that SEPTIC builds for the \vwebapp{}s
that use OOP, SEPTIC can defend against only four variants of SQLi attacks in
our dataset.
All 4 SQLi attacks that SEPTIC can defend against reside in Wordpress.
Wordpress does not use the encapsulation concept in its database API, and its
modules provide SQL queries as function arguments; consequently, SEPTIC can
correctly create its mapping.
\vP{} overcomes this problem by utilizing a static analysis that identifies
the database API in the \vwebapp{}, which helps \vP{} to correctly determine the
function that interacts with the database.
Merlo et al. ~\cite{merlo2007automated} proposed a two step approach.
First, it intercepts every function call to \texttt{mysql\_query} and records a
profile for benign issued SQL queries.
The profile is a mapping between the issued SQL query and the PHP function that
calls the function \texttt{mysql\_query}.
During enforcement,~\cite{merlo2007automated} looks for the received SQL query
in its mapping profile and if the query does not syntactically match with any
recorded query for the PHP function, \cite{merlo2007automated} blocks the
query.
The proposed approach in \cite{merlo2007automated} maps all the SQL queries to
the internal functions in the database API instead of the appropriate function
that uses the database API for communicating with the database.
Besides that, enforcing a strict comparison of the parse tree limits the
functionality of the \vwebapp{} for generating dynamic SQL queries.
Table~\ref{tab:compare} shows that the proposed approach in
~\cite{merlo2007automated} cannot protect \vwebapp{}s against any of the SQLi
attacks in our dataset.
\textit{Hybrid Analysis}: 
Amnesia~\cite{halfond2005amnesia} builds a model of benign queries in Java
\vwebapp{}s statically. 
At runtime, Amnesia checks the queries passed to the database against the built
model. 
Amnesia highly depends on the benign queries that are built during the static
analysis, which leads to a high number of false positives when applied to
programs that generate SQL queries dynamically.
SQLRand~\cite{boyd2004sqlrand} proposed a randomization technique for
randomizing queries in \vwebapp{}s. 
SQLRand randomizes the SQL queries in the \vwebapp{} and uses an intermediary
proxy for de-randomizing before sending the queries to the database.
Since \vwebapp{}s generate SQL queries dynamically, randomizing the queries
using SQLRand is a challenging task.
Using an intermediate proxy introduces overwhelming overhead to \vwebapp{}
performance~\cite{halfond2006classification, su2006essence}.
Besides that, since there is one static key that modifies SQL keywords, the
knowledge of new SQL keywords can compromise the security of
SQLRand~\cite{buehrer2005using}.

\section{System Overview}\label{sec:sysoverview}
In this section we explain how \vP{} records benign SQL queries and limits
the access of functions in a \vwebapp{} to the database.
Figure~\ref{fig:sys} shows an overview of how \vP{} defends \vwebapp{}s against
SQLi attacks.
Specifically, \vP{} records a profile by observing benign issued queries by a
\vwebapp{}.
\vP{} then enforces the profile from inside the database for every query that
the \vwebapp{} sends to the database.
In step \vsone{}, \vP{} performs a static analysis over the \vwebapp{} to
identify the database procedures that are used across the \vwebapp{}'s scripts.
This analysis is done once per \vwebapp{} and \vP{} uses this information
during training and enforcement of the profile.
In step \vstwo{}, \vP{} is in the training mode and records the benign issued
SQL queries by the \vwebapp{}. 
\vP{} can use benign browsing traces or the \vwebapp{}'s unit tests in its
training.
\vP{} creates a mapping between the benign SQL queries that MySQL receives and
the functions in the \vwebapp{} that used the \vdbal{} to issue the query to
the database.

%
In Step \vsthree{}, \vP{} leverages the information from the first two steps to
assemble a trusted database-access profile.
The profile is a set of allowed tables, SQL functions, and type of SQL queries
that each function in the \vwebapp{} can issue.
At the end of the third step, \vP{} acquires the necessary information to
protect the \vwebapp{} from SQLi attacks. 
In step \vsfour{}, \vP{} protects the running \vwebapp{} against unauthorized
database access by filtering access to the database according to the trusted
profile generated in step \vsthree{}.
The modified database extension (e.g., PDO in PHP) appends the execution
information (i.e., \vcs{}) at the end of each SQL query as a comment before
sending it to MySQL.
Prior to the execution of each SQL query, \vP{} extracts the appended execution
information from the SQL query and identifies the function that communicate
with the database using the \vdbal{}.
\vP{} checks the query against the profile that corresponds to the function
that issued the query. 
Finally, if the SQL query matches the profile, MySQL executes the query and
returns the results.

\subsection{Static Analysis of \vWebapp{}s} 

\lstset{
  basicstyle=\ttfamily,
  columns=fullflexible,
  keepspaces=true,
  numbers=left,
}

\definecolor{dkgreen}{rgb}{0,0.6,0}
\definecolor{dkblue}{rgb}{0,0,0.6}
\definecolor{dkyellow}{cmyk}{0,0,0.8,0.3}
\lstset{
  language        = php,
  keywordstyle    = \color{dkblue},
  stringstyle     = \color{ACMDarkBlue},
  identifierstyle = \color{ACMBlue},
  commentstyle    = \color{gray},
  emph            =[1]{php},
  emphstyle       =[1]\color{black},
  emph            =[2]{if,and,or,else},
  emphstyle       =[2]\color{Periwinkle},
  escapechar=|,
  showstringspaces=false}
\begin{figure}[!b]
	\begin{adjustbox}{minipage=\columnwidth,scale=0.7}
\begin{subfigure}[b]{\columnwidth}
\begin{lstlisting}
$id = $_GET["id"]
function get_public_info{
include dirname(__FILE__)."/db/database.php";
$users = executeQuery("public_info", $id);|\label{line:callproc}|
...
}
get_public_info();
\end{lstlisting}
	\caption{\texttt{get\_public\_info.php}}
	\label{fig:code_b}
\end{subfigure}
\begin{subfigure}[b]{\columnwidth}
\begin{lstlisting}
class DatabaseConnectionmysqli |\label{line:startclass}|
	extends mysqli {
  private $query;
  function __construct(){
      parent::__construct("localhost","admin","admin","mysqldb");
  }
  public function setQuery( $query ){|\label{line:startsetquery}|
      $this->query = $query; |\label{line:bodysetquery}|
      ...
  }
  public function execute(){|\label{linestartexecute}|
      return parent::query($this->query);|\label{line:bodyexecute}|
  }
  public function multi_execute(){|\label{linestartmexecute}|
      $result = parent::multi_query($this->query);|\label{line:bodymexecute}|
      ...
  }
}
function executeQuery( $tbl, $arg ) { |\label{line:startproc}|
      $query = "SELECT * FROM ".$tbl." WHERE id > ".$arg;
      $classname = "DatabaseConnection".$this->getDriver();|\label{line:clsname}|
      return new $classname()->setQuery($query)->multi_execute();|\label{line:instnew}|
}
\end{lstlisting}
\caption{\texttt{/db/database.php}}
\label{fig:code_c}
\end{subfigure}
	\end{adjustbox}	
	\caption{Illustrative PHP code snippets demonstrating dynamic inputs to \textit{new} keyword}
\label{fig:codesample}
\end{figure}

The \vwebapp{} \vdbal{} provides a unified interface to interact with
different databases.
In step \vsone{}, \vP{} identifies the \vdbal{} by statically analyzing the
\vwebapp{}. 
To this end, \vP{} creates a class dependency graph (CDG). 
The CDG is a directed graph $CDG=(V,E)$, where the vertices ($V$) are classes
and interfaces in the \vwebapp{}. 
An edge $e_{1,2} \in E$ is drawn between $v_1 \in V$ and $v_2 \in V$ if $v_1$ extends class
$v_2$, implements interface $v_2$ . 
After creating the CDG, \vP{} extracts the list of classes and interfaces in
the \vwebapp{} that extend database APIs (e.g., PDO in PHP).
To do so, we manually identify database extension classes (e.g.,
\texttt{mysqli} in PHP). 
Afterwards, \vP{} iterates over the vertices of the CDG and checks whether a
vertex is connected to the database API.
If a vertex is connected to the database API, \vP{} adds it to the \vdbal{}.
\vP{} also adds classes to the \vdbal{} if their methods initialize an instance
of database API in PHP (e.g., \texttt{mysqli\_init}).
At the end of this iteration, \vP{} possess a list of all classes and
interfaces in the \vwebapp{} that extends the database API.
Besides the object-oriented design of \vdbapi{}s in \vwebapp{}s, operations on
databases (e.g., \texttt{SELECT} operation) also have
procedures~\cite{drupaldapi}. 
Database procedures handle creation of objects from \vdbapi{} and setting
correct parameters for modules in the \vwebapp{}.
A database procedure returns an object from a sub-type of a database API.
\vP{} analyzes the body of the functions and procedures in the \vwebapp{} for
the returned objects.
If the returned object is from a sub-type of a \vdbapi{} in the \vwebapp{},
then \vP{} considers it as a database procedure.
At the end of this step, \vP{} extracts information regarding the database API
as well as database procedures. 
This step is necessary for \vP{} to find the function that used the \vdbal{}
for communicating with the database during training and enforcing of the
profile.
Figure~\ref{fig:code_c} shows a snippet of PHP code from a class  that extends
the database API \texttt{mysqli}.
There is also a database procedure called the \texttt{executeQuery} in
Figure~\ref{fig:code_c} that return an object from
\texttt{DatabaseConnectionmysqli} that is a subclass of \texttt{mysqli}. 
Figure~\ref{fig:code_b} shows another snippet of code implementing a function
called \texttt{get\_public\_info} that uses \texttt{executeQuery} to retrieve
data from the database.
\vP{} identifies \texttt{DatabaseConnectionmysqli} as a subclass of
\texttt{mysqli} and \texttt{executeQuery} as a database procedure.

\subsection{Collecting information regarding database access in the \vwebapp{}}
In step \vstwo{}, we train \vP{} using benign traces or unit tests to learn
benign SQL queries.
Step \vstwo{} consists of  two components that work together to create a
mapping between the received SQL query in MySQL and the function that composed
the SQL query.
The first component, appends the execution information at the end of each SQL
query before sending it to the database.
The execution information includes the \vcs{} in the \vwebapp{} that led to
sending a SQL query to the database using database extensions (e.g.,
\texttt{PDO} or \texttt{mysqli} in PHP).
The second part, a MySQL plugin, intercepts the execution of the incoming SQL
queries to MySQL.
When MySQL receives a SQL query through benign traces or unit tests, \vP{}
records the SQL query that MySQL receives including the execution information
appended to the SQL query.
Since \vP{} has access to the parse tree of the SQL query, \vP{} traverses the
parse tree and records information regarding the type of nodes in the parse
tree of the SQL query.
\vP{} also logs the list of tables that the SQL query accesses, as well as the
type of operation (e.g, \texttt{SELECT} operation) in the SQL query.

\subsection{Creating the profile}\label{subsec:sys:RI} 
\vP{} in step \vsthree{}, leverages the access logs collected from  benign SQL
queries in step \vstwo{} and generates a profile that defines the access to the
database for each function in the \vwebapp{} that interacted with the database.
Particularly, the profile contains a set of query descriptors for each function
in the \vwebapp{}.
A query descriptor comprises four components.
Each component specifies a different aspect of the database access, that we
explain below.
\begin{itemize}
	\item 
		\textit{\textbf{Operation:}} denotes the type of operation in
		the SQL query. The operation can be \texttt{SELECT, INSERT,
		UPDATE, DELETE}, etc.~\cite{mysql}.
		The profile records the type of operation in each SQL query. 
		Enforcing the operation type removes the possibility of a SQLi
		attack performing a different operation. 
		For instance, when the profile only specifies a \texttt{SELECT}
		operation, the SQL query cannot perform an \texttt{INSERT} SQL
		query.
	\item \textit{\textbf{Table:}} determines the tables that the SQL query
		can operate on.  
		Restricting the tables used in a SQL query prevents  an
		attacker from executing a SQL query on a different table.  
	\item 
		\textit{\textbf{Logical Operator:}} indicates the logical
		operators~\cite{mysql} used in the SQL query. 
		Logical operators limit the ability of an attacker to use a
		tautology attack in a SQL query for extracting data from a
		table.
	\item
		\textit{\textbf{SQL function:}} determines the list of
		functions that the query uses. 
		The component also records the type of arguments that are
		passed to each function. 
		The list of functions restricts the attacker to use only the
		functions that are recorded during the training. 
		This limits the attacker's capability to use alternate encoding
		and stored-procedures attacks against the database.

\end{itemize}
At the end of step \vsthree{}, \vP{} acquired a set of query descriptors for
each function in the \vwebapp{} that issued a SQL query based on the training
data that was obtained in step \vstwo{}.
\subsection{Protecting the \vWebapp{}} 
In the last step, \vP{} is in enforcement mode and uses the profile created in
step \vsthree{} to restrict access to the database for each function in the
\vwebapp{}.
When the database receives a SQL query, \vP{} extracts information regarding
the type of operation, table accesses, and parse tree of the received SQL
query.
Subsequently, \vP{} extracts the function that issued the SQL query from the
execution information appended to the incoming SQL query.
Afterwards, \vP{} looks up in the profile and retrieves query descriptors
associated with the function that composed and issued the SQL query.
For each query descriptor associated with function, \vP{} compares each
component of the query descriptor with the obtained information from the
received SQL query.
First, \vP{} checks whether the type of operation in the received SQL query and
in the query descriptor is the same or not.
Second, \vP{} examines the list of tables in the received SQL query.
The list of table in the received SQL query must be a subset of the list of
tables in the query descriptor.
For the logical operators, \vP{} checks whether the logical operators in the
SQL query that MySQL received is subset of the logical operators in the query
descriptor.
Finally, \vP{} inspects the functions used in the received SQL query as well as
the type of arguments.
The functions and the type of arguments must be in the recorded query
descriptor.
\vP{} takes a conservative approach and allows the database to execute the SQL
query only if all four components of a query descriptor associated with the
function authorize the SQL query. 

\section{Implementation}\label{sec:implementation} 
\lstset{
  basicstyle=\fontsize{8}{11}\selectfont\ttfamily,
  columns=fullflexible,
  keepspaces=true,
  numbers=left,
}

\definecolor{dkgreen}{rgb}{0,0.6,0}
\definecolor{dkblue}{rgb}{0,0,0.6}
\definecolor{dkyellow}{cmyk}{0,0,0.8,0.3}
\lstset{
  language        = php,
  keywordstyle    = \color{dkblue},
  stringstyle     = \color{ACMDarkBlue},
  identifierstyle = \color{ACMBlue},
  commentstyle    = \color{gray},
  emph            =[1]{php},
  emphstyle       =[1]\color{black},
  emph            =[2]{if,and,or,else},
  emphstyle       =[2]\color{Periwinkle},
  escapechar=|,
  showstringspaces=false}

\begin{figure*}[!t]
\begin{adjustbox}{width=0.9\textwidth,keepaspectratio}
\begin{lstlisting}
SELECT * FROM public_info where id > 0 # mysqli::multi_query@DatabaseConnectionmysqli::multi_execute@executeQuery@get_public_info
FIELD@FUNC:>@2@FIELD@LITERAL # recorded info regarding the nodes in the SQL query
public_info@0 # recorded info regarding the table and operation type of the SQL query
\end{lstlisting} 
\end{adjustbox}	
	\caption{\texttt{recorded information regarding the
	execution of function \texttt{get\_public\_info}}} 
\label{fig:profile}

\end{figure*}

In this section we elaborate on the implementation challenges that needed to be
addressed build \vP{}.
First, we explain how \vP{} statically analyzes PHP \vwebapp{}s to identify the
\vdbal{}.
Afterwards, we describe how \vP{} uses the MySQL plugin API to record the SQL
queries that the database receives. 
We explain how \vP{} creates a precise profile for each PHP function based on
the SQL queries issued to the database.
Finally, we describe \vP{}'s approach for using a MySQL plugin API to restrict
database accesses.  
\subsection{Static Analysis of \vwebapp{}s } 

In step \vsone{}, \vP{} analyzes the \vwebapp{} to determine the \vdbapi{} and
database interfaces across the PHP scripts in a \vwebapp{}. 
\vP{} performs a flow-insensitive analysis, which focuses on
finding \vdbapi{}, interfaces, and procedures. 

\vP{} identifies all PHP files in the \vwebapp{}, using \texttt{libmagic}. 
We use php-parser~\cite{slizov2018} to parse each PHP script into an abstract
syntax tree (AST). 
\vP{} identifies classes, interfaces, and abstract definitions by scanning AST
nodes that represent their corresponding definitions. 
\vP{} examines interface and class definitions across the PHP \vwebapp{} to
reason about the dependencies between classes and interfaces,  
During analysis, \vP{} creates a class dependency graph (CDG) and draws an edge
between interfaces and classes when: 1) An interface extends another interface.
2) A class implements an interface. 3) A class extends another class.
After creating the CDG, the static analyzer (\vSA{}) iterates over the nodes of the CDG
to identify classes and interfaces that facilitate communication between the
PHP \vwebapp{} and the database.
To accomplish this, \vP{} starts with the \texttt{PDO} and \texttt{mysqli} classes;
two of the most popular database extensions in PHP.
\vP{} creates a list of classes and interfaces that share an edge with \texttt{PDO} or
\texttt{mysqli} classes in the CDG.
For example, after creating the CDG for the code in Figure~\ref{fig:code_c},
\vSA{} identifies \texttt{DatabaseConnectionmysqli} as a subclass of \texttt{mysqli}.
\vSA{} must identify database procedures as well.
\vSA{} decides whether a procedure is a database procedure or not by analyzing
the type of object it returns.
If a procedure returns an object from a subclass of the database API, \vSA{}
marks that function as a database procedure.
For determining the object type that a function returns, \vSA{} analyzes the
AST node of the return statement.
There are two cases that \vSA{} is interested to follow:
\begin{itemize}
	\item{\textbf{Instantiating an object using the \texttt{new} keyword:}}
		If the function is instantiating an object using the
		\texttt{new} keyword in the return statement, \vSA{} analyzes
		the argument that is passed to the \texttt{new} keyword. If the
		argument is the name of a subclass of a database API, \vSA{}
		marks the function as a database procedure. If the argument is
		a variable, \vSA{} performs a lightweight static analysis as a
		limited form of constant folding over strings that compose the
		value. \vSA{} marks the function as a database procedure, if
		the resolved value is a subclass of database API.
	\item{\textbf{Variable:}} If the function returns a variable,
		\vSA{} iterates backward on the AST to the last assignment of the
		variable and checks whether the assignment is a class
		instantiation or not. If it is a class instantiation, \vSA{}
		tries to resolve the type of instantiated object as described
		above.
\end{itemize}	
As discussed in \S~\ref{subsec:back:php}, PHP \vwebapp{}s often use variables
as an argument for creating objects from classes using the \texttt{new}
keyword.
During analysis, \vSA{} keeps track of arguments passed to \texttt{new} in
PHP scripts using a string representation.  
\subsubsection{String Representation} 
\vSA{} encounters strings when handling variable assignments and constant
definitions. 
Strings can be a mixture of literal components, function return
values, and variables. 
When \vSA{} iterates over an assignment node in the AST, it records a set of
information from the assignment node in a hash table.
\vSA{} keeps track of  the name of the variable and the components on the right
side of the assignment. 
\vSA{} also records the name of the function or the name of the class and
method that the assignment statement occurs in.
For example, in Figure~\ref{fig:code_c} at Line~\ref{line:clsname}, the
function \texttt{executeQuery} has an assignment statement. The right side of
the assignment concatenates a constant string and a return value from a
function.
\vSA{} records the name of the variable on the left side of the assignment as
well as the value of the constant string and the return value from the
function.
\vSA{} also records the type of operation on the right side (as discussed next,
it is a concatenation operation).
\vSA{} implements common string operations to resolve the value of the
assignment.

\subsubsection{String Operations} \vP{} manages frequent string-related
operations.

\textbf{Variables:} 
The argument passed to \texttt{new} can contain variables defined in the
script. 
\vSA{} keeps track of variable definition in the scope of script, class, or
functions. 
When there is a variable assignment, \vSA{} creates an object for the variable
and its value. 
\textbf{Concatenation:} 
In PHP, strings can be constructed by joining multiple components with the
\texttt{.} and \texttt{.=} operators. 
\vSA{} handles string concatenation by creating an object for concatenation and
adds components that exist in the concatenation statement.
\subsubsection{Identifying Database Procedures}
To identify database procedures, \vSA{} iterates over the assignments and
resolves the value of variables in the strings by looking for variables in the
same class and function. 
If there is a variable without a value, \vSA{} represents the value as a
regular expression \texttt{.*} wildcard. 
\vSA{} looks for a match between the generated regular expression and the list
of database API subclasses. 
For example, in Figure~\ref{fig:code_c}, line~\ref{line:clsname}, \vSA{} cannot
determine the return value of \textit{\$this->getDriver}. 
Instead, \vSA{} represents the value as a \texttt{.*} wildcard. 
\vSA{} searches the list of database API subclasses for a class that matches
the regular expression \texttt{DatabaseConnection*}, and finds such a class
named \texttt{DatabaseConnectionmysqli}. 
\vSA{} marks \textit{executeQuery} as a database procedure. 
At the end of this step, \vSA{} has a list of \vdbal{} classes, interfaces, and
procedures.
\subsection{Profile Data Collection}
\label{subsec:map}
This step trains \vP{} to create a mapping between issued SQL queries and the
\vwebapp{}'s function that relied on the \vdbal{} to issue the SQL query.
The collected information in this step is necessary for generating query
descriptors in step \vsthree{}.
As described in \S~\ref{sec:sysoverview}, the information collected for
each SQL query contains the operation, the access tables, the logical
operators, the SQL functions that the query used, and the type of arguments in
each SQL function.
\rasoul{ It is not query descriptor yet, step three will create query
descriptors using this information }

\subsubsection{Attaching a PHP call stack: } \label{subsec:avcs}
When MySQL receives a SQL query, \vP{} must infer which PHP function actually issued
the SQL query. 
To achieve this, we modified the source code of the MySQL driver for the
\texttt{PDO} and \texttt{mysqli} extensions.
This modification appends the PHP \vcs{} at the end of the query as a comment
before sending it to the database. 
To access the PHP \vcs{}, we use the Zend framework's built-in function called 
\texttt{zend\_fetch\_debug\_backtrace}. 
Zend keeps the information regarding the \vcs{} for the executing PHP script\@. 
This information includes the functions, class, their respective arguments, the
file, and the line number that issued the call. 
The modified database extension (\vDE{}) extracts the PHP call
stack and appends it as a comment to the end of the SQL query. 
\subsubsection{Extracting information from the parse tree: } 
Recorder plugin (\vPR{}) acts as a post-parse MySQL plugin.
\vPR{} has access to various information regarding the parsed SQL query in
MySQL: the type of operation (e.g. \texttt{SELECT} operation, etc.), the name
of the table, and the parse tree of the SQL query. 
MySQL provides a parse tree visitor function that \vPR{} uses to access the
parse tree of SQL queries.
However, MySQL only allows plugins to access literal values of the query, such as user inputs in the
parse tree.  
Because \vP{} needs more information regarding the parsed SQL query, we
modified the source code of MySQL-server so that the plugin can access
non-Literal values as well.
When MySQL invokes \vPR{}, \vPR{} records the SQL query that MySQL receives.
Afterwards, \vPR{} iterates over the parse tree of the SQL query and records
the type of each node.
If the node represents a SQL function in the SQL query, \vPR{} also
records the number of arguments used in the SQL function.
The node that represents the SQL function in the SQL query also holds the
number of arguments used in the SQL function.
Afterwards, \vPR{} records the type of arguments passed to the SQL function as
they appear in the parse tree of the SQL query.
Lastly, \vPR{} logs the table and the type of operation for the SQL query that
MySQL received.
In MySQL the information regarding the type of operation for a SQL query is
shown as a number.
Hence, \vPR{} logs the type of operation for a SQL query as an encoded number
in the profile.
Figure~\ref{fig:profile} shows the recorded information in the profile, when
function \texttt{get\_public\_info} executes.
At the end of step \vstwo{}, \vP{} has detailed information on the received
SQL queries for training.
\subsection{Creating the Profile} \label{subsec:definerule}
In step \vsthree{}, profile generator (\vPG{}) creates a profile for each PHP
function in the \vwebapp{} that accesses the database. 
\vPG{} relies on the training data from step \vstwo{} as input.
\vPG{} reads the recorded information from step \vstwo{}.
As shown in Figure~\ref{fig:profile}, the first line is the SQL query including
the PHP \vcs{}.
Using the list created in step \vsone{}, \vPG{} must infer which PHP used the
\vdbal{} to send the SQL query to the database.
This is a difficult problem, because the last function on the call stack might
be a helper function that issues all queries for the application (and, in fact,
this is how modern real-world PHP applications such as Wordpress and Joomla are
written).
\vPG{} iterates over the stack of functions in the PHP \vcs{} and checks
whether the function or the method was recognized as a database procedure or
database API method in step \vsone{}.
\vPG{} iterates over the stack starting from the last call in PHP \vcs{} until
a function is not a database procedure or database API method. 
\vPG{} identifies this function as the function that created the database
query.
As an example, the Line~1 in Figure~\ref{fig:profile} shows the SQL query that
MySQL receives including the PHP \vcs{}.
\vPG{} detects \texttt{mysqli} as a database extension in PHP and
\texttt{DatabaseConnectionmysqli} as a class that extends \texttt{mysqli}.
Then, \vPG{} visits the next function \texttt{executeQuery}, which was
identified as a database procedure in step \vsone{}.
The next function in the PHP \vcs{} is \texttt{get\_public\_info}. 
\texttt{get\_public\_info} is not in the list of database procedures from step
\vsone{}, therefore \vPG{} identifies it as the PHP function that used \vdbal{}
to send the SQL query to the database.
\vPG{} will then update \texttt{get\_public\_info}'s query descriptor.
Afterwards, \vPG{} iterates over the nodes of the SQL query's parse tree and
extracts all the logical operators.
If all the logical operators are the same, \vPG{} updates the \textit{cond}
with the respective value.
If both logical operators (i.e, both \textit{OR} and \textit{AND}) are in the
nodes of SQL query's parse tree, \vPG{} sets \textit{cond} to
\textit{"Both"}.
If there is no logical operators in the SQL query, \vPG{} sets \textit{cond} to
\textit{"None"}.
Based on Figure~\ref{fig:profile}, \vPG{} specifies that
\texttt{get\_public\_info} does not use any logical operators in its SQL query.
\vPG{} iterates over the list of nodes from the parsed tree of the SQL query
and extracts the name of the used functions in the SQL query as well as their
respective arguments. 
Since the number of arguments passed to the SQL function can be variable,
\vPG{} does not record each argument's type.
Instead, \vPG{} summarizes the types of arguments that a SQL function relies on.
There are multiple types of functions in MySQL such as numeric, string,
comparison, and date function.
All of the aforementioned types of SQL functions except the comparison type either
receive less or equal to two arguments or modifies the content of the first
argument passed to the function.
Comparison functions in MySQL (e.g., \texttt{<}, \texttt{IN}, etc.) compare a
single argument to a variable sized argument array.
Moreover, the single argument appears as the first argument in the SQL
comparison functions.
Owing to this, \vPG{} records the type of the first argument passed to a SQL
function separately.
If the argument is a table column, \vPG{} records it as a \texttt{FIELD}
argument, otherwise \vPG{} records it as a \texttt{LITERAL} argument.
Afterwards, \vPG{} iterates over the rest of the arguments passed to the SQL
function.
If the type of all the other arguments are the same type (i.e., \texttt{FIELD}
or \texttt{LITERAL}), then \vPG{} records the value of the respective type in
the profile.
Otherwise \vPG{} sets the type as \textit{var}.
For instance, based on Figure~\ref{fig:profile}, \vPG{} specifies that function
\texttt{get\_public\_info} used function \texttt{">"}, that the first argument
is a table column and the second argument is a \texttt{LITERAL}.
Lastly, \vPG{} reads the information about the name of the table and the type of SQL
query. 
For instance, based on line 3 in Figure~\ref{fig:profile}, \vPG{} deduces that
function \texttt{get\_public\_info} accesses the  table \texttt{public\_info}
using a \texttt{0}-type SQL query (i.e., \texttt{SELECT} SQL query).
At the end of step \vsthree{}, \vPG{} has a set of query descriptors for each
PHP function in the \vwebapp{} that issued a SQL query during training in step
\vstwo{}
\subsection{Protecting the \vwebapp{}}
In step \vsfour{}, the enforcer plugin (\vPE{}) is on enforcement mode.
\vPE{} uses the profile that was generated in step \vsthree{} and protects the
database from queries that deviate from the profile. 
Similar to \vPG{}, \vPE{} is implemented as a postplugin, which gives it access
to the parse tree of the received SQL query.
\vPE{} also uses the same PHP database extensions as described in
\S~\ref{subsec:avcs}.
\vPE{} reads the profile for each PHP function and uses it to analyze the
received queries. 
After receiving a query, MySQL parses the SQL query and calls \vPE{}. 
\vPE{} locates the \vcs{} and extracts the PHP function that issued the query
with the same approach described in \S~\ref{subsec:definerule}. 
Afterwards, \vPE{} finds the query descriptors in the profile associated with
the PHP function.
\vPE{}  checks the query against all four components of each query descriptor
found for the PHP function.
For operation type, \vPE{} checks whether the received SQL query has the same
operation type as it is recorded in the profile.
\vPE{} also examines that the list of tables accessed for the received SQL
query is a subset of table access listed in the query descriptor.
The logical operators used in the received SQL query must be a subset of the
logical operators in the query descriptor.
Finally, the received SQL query can only use a subset of functions listed in
the query descriptor.
\vPE{} also checks whether the arguments passed to each function has the same
type as it is recorded in the query descriptor.
Only if the SQL query matches with all four components of at least one query
descriptor in the profile, \vPE{} allows MySQL to execute the SQL query and
return the results.
Otherwise \vPE{} returns \texttt{False} to MySQL-server, aborting execution of
the query and returning an error to the \vwebapp{}, thus preventing a
potentially malicious attacker-controlled SQL query from executing.
%

\section{Evaluation}\label{sec:eval} 

We assessed the ability of \vP{} to prevent SQLi attacks on a set of
popular PHP \vwebapp{}s. 
We also examined \vP{}'s false positive rate during the benign browsing
of the \vwebapp{}. 
Additionally, we evaluated the performance overhead of step \vsthree{} for the
benign browsing. 
For our evaluation, we answer the following research questions: 

\noindent 
	\begin{tabularx}{\columnwidth}{l@{\hspace{3pt}}X} 
	\noindent 
	\small
	\textbf{RQ1} &\small How precise is \vP{}'s static analysis?
	\\
	\small
	\textbf{RQ2} &\small Is \vP{} effective against real world SQLi
		vulnerabilities in popular \vwebapp{}s?
	\\ 
	\small \textbf{RQ3} &\small How practical is \vP{} regarding
		performance overhead and false positives?

\end{tabularx}
\subsection{Evaluation Strategy} \label{subsec:strategy}
In our evaluation, we performed our static analysis once for each \vwebapp{} in
Section~\ref{subsec:evaldataset}.
We evaluated the \vdbal{} resolved by our static analysis in \texttt{RQ1}.
Then we leveraged our \vdbal{} to answer \texttt{RQ2} and \texttt{RQ3}.
We trained and built the profile for \vP{} using the official unit tests of
each \vwebapp{} once and used the generated profile for the experiments to
answer \texttt{RQ2} and \texttt{RQ3}.
The official unit tests examine the correctness of functions in the \vwebapp{}
by executing test-inputs and verifying their results.
The advantage of unit tests over web crawlers is that there is no need for
manual intervention of administrators, specifically for providing semantically
correct inputs for each form in \vwebapp{}s.
A \vwebapp{}'s unit tests are specifically tailored to its implementation and
therefore are likely to achieve higher code coverage.
\begin{figure}[h!]
	\includegraphics[width=0.7\columnwidth]{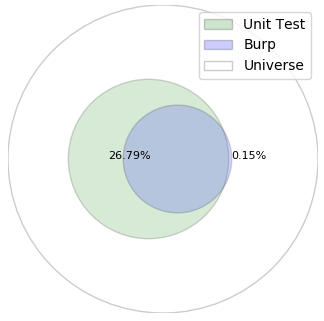}
	\caption{The line coverage for unit tests and Burp suite on Drupal 7.0}
	\label{fig:coverage}
\end{figure}
Figure~\ref{fig:coverage} shows that Drupal's unit tests achieve higher line
coverage compared to Burp suite and also covers almost all the lines that Burp
Suite~\cite{burp} covered.
However, alternative approaches such as web crawlers can also be used for
training \vP{}.

\subsection{Evaluation Dataset} \label{subsec:evaldataset} 
We evaluated \vP{} on the four most popular PHP \vwebapp{}s, Wordpress, Joomla,
Drupal, and Magento.
According to W3Techs, these \vwebapp{}s hold \vcmsstat{} of the market share
among all existing content management systems (CMS) and power \vwebstat{} of
all the live websites on the Internet combined~\cite{w3marketshare}.  
Administrators install plugins and additional components to customize the
\vwebapp{} and extend its functionality. 
To reflect this behavior in our evaluation, we also evaluate \vP{} on plugins.
We installed four vulnerable Wordpress plugins called \textit{Easy-Modal},
\textit{Polls}, \textit{Form-maker}, and \textit{Autosuggest}.
We also installed three vulnerable plugins in Joomla named \textit{jsJobs},
\textit{JE photo gallery}, and \textit{QuickContact}.
To assess the defensive capability of \vP{}, we selected recent versions of the
\vwebapp{}s and plugins that contain known SQLi vulnerabilities. 
We also considered the type of SQLi vulnerability in our dataset to include all
types of SQLi exploits for a comprehensive evaluation. 
We collected a total of \vdataset{} SQLi vulnerabilities in different
\vwebapp{}s and plugins.

\subsection{Resolving The Database Access Layer (RQ1)}
In step \vsone{}, \vP{} scans the PHP \vwebapp{} to identify the \vdbal{} that
is used to communicate with the database.
Step \vsone{} is a crucial step to identify the correct function in the PHP
\vcs{} that relies on the \vdbal{} for interacting with the database.
Table~\ref{tab:satab} presents the resolved \vdbal{} statistics.
The \textit{resolved subclasses} column specifies the number of classes that
extends the database API in PHP.
The \textit{resolved database procedures} column presents the number of
functions that returns an object from a subclass of the database API.
Since there is no ground truth for the \vdbal{} in the \vwebapp{}s, we manually
analyze the output of \vSA{} for true positives.
Subclasses of database APIs in the PHP \vwebapp{}s also implement interfaces
to facilitate actions such as iterating over elements in the object and
counting elements.
For instance, Drupal implements \texttt{Iterator} and \texttt{Countable} so
that the PHP script can iterate over or count the number of records that the
database returns to the PHP script.
Since Drupal implements \texttt{Countable} and \texttt{Iterator} in the
subclasses of database API,\vSA{} adds these two interfaces to the \vdbal{}.
As shown in Table~\ref{tab:satab}, the only false positives we observed during
our evaluation are caused by the \texttt{Iterator} and \texttt{Countable}
interfaces.
All the \vwebapp{}s in our dataset except for Wordpress, use encapsulation in their
database API subclasses and database procedures that show the necessity of
identifying the \vdbal{} for creating a profile.
Without identifying the \vdbal{}, \vP{} would operate similar to SEPTIC and
map the received queries to a single identifier.

\begin{table}[!h]
\resizebox{\columnwidth}{!}{
\begin{tabular}{lcc}
	\vWebapp{}  & Resolved subclasses (FP) & Resolved database procedure\\\hline
	Wordpress 4.7 	& 1 		& - 	\\
	Drupal 7.0 	& 44 (2) 	& 38	\\
	Joomla 3.7 	& 30 (0) 	& -	\\
	Joomla 3.8 	& 30 (0) 	& -	\\
	Mangeto 2.3.0 	& 15 (0) 	& -	\\\hline
\end{tabular}
}
	\caption{Resolved \vdbal{}}
\label{tab:satab}
\end{table}

\subsection{Defensive Capabilities (RQ2)}\label{rq2}
We assessed the defense capabilities of \vP{} against \vdataset{} SQLi
vulnerabilities listed in Table~\ref{tab:defense}. 
We built and deployed five Docker containers that run a vulnerable version of a
\vwebapp{} and a plugin. 
We exploit the vulnerabilities using exploits from Metasploit
Framework~\cite{metasploit}, exploit-db~\cite{exploitdb}, and
sqlmap~\cite{sqlmap}. 
We consider an attack successful if an attacker can inject malicious SQL code
into the generated query in the \vwebapp{} and the database executes the
malicious SQL query.
For this evaluation we used the results of our static analysis in \texttt{RQ1}.
We trained \vP{} using the official unit tests of \vwebapp{}s in their
respective repositories.
After creating the profile, we configured \vP{} in the enforcement mode and
assess whether the exploits in exploit-db and Metasploit Framework are
successful or not.
Adversaries are not limited to use exploits in our evaluation and can craft
their SQL queries to circumvent \vP{}. 
To evaluate the potential of such attacks, we also used sqlmap~\cite{sqlmap} to
generate various exploits for the vulnerabilities listed in
Table~\ref{tab:defense}.

\newcommand\vAapp{Wordpress 4.7}
\newcommand\vBapp{Drupal 7}
\newcommand\vCapp{Joomla 3.7}
\newcommand\vDapp{Joomla 3.8.3}
\newcommand\vEapp{magento 2.3.0}

\newcommand\vAplugin{Easy-Modal}
\newcommand\vBplugin{Polls}
\newcommand\vCplugin{Form-maker}
\newcommand\vDplugin{Autosuggest}
\newcommand\vEplugin{\vCore{}}
\newcommand\vFplugin{\vCore{}}
\newcommand\vGplugin{jsJobs}
\newcommand\vHplugin{JE photo gallery}
\newcommand\vIplugin{QuickContact}
\newcommand\vJplugin{\vCore{}}
\newcommand\vKplugin{\vCore{}}

\newcommand\vAvuln{\href{https://nvd.nist.gov/vuln/detail/CVE-2017-12946}{CVE-2017-12946}}
\newcommand\vBvuln{\href{https://www.exploit-db.com/exploits/44229}{polls-widget 1.2.4}}
\newcommand\vCvuln{\href{https://nvd.nist.gov/vuln/detail/CVE-2019-10866}{CVE-2019-10866}}
\newcommand\vDvuln{\href{https://wpvulndb.com/vulnerabilities/9188}{WPVDB-9188}}
\newcommand\vEvuln{\href{https://nvd.nist.gov/vuln/detail/CVE-2014-3704}{CVE-2014-3704} \vCore{}}
\newcommand\vFvuln{\href{https://nvd.nist.gov/vuln/detail/CVE-2017-8917}{CVE-2017-8917} \vCore{}}
\newcommand\vGvuln{\href{https://www.exploit-db.com/exploits/47249}{com\_jsjobs 1.2.5}}
\newcommand\vHvuln{\href{https://www.exploit-db.com/exploits/45930}{com\_jephotogallery 1.1}}
\newcommand\vIvuln{\href{https://nvd.nist.gov/vuln/detail/CVE-2018-5983}{CVE-2018-5983}}
\newcommand\vJvuln{\href{https://nvd.nist.gov/vuln/detail/CVE-2018-17385}{CVE-2018-17385} \vCore{}}
\newcommand\vKvuln{\href{https://nvd.nist.gov/vuln/detail/CVE-2019-7139}{CVE-2019-7139} \vCore{}}

\newcommand\vAtype{Taut., Infer., Alt. Encoding } 
\newcommand\vBtype{Taut., Infer., Alt. Encoding} 
\newcommand\vCtype{Infer.} 
\newcommand\vDtype{Taut., Infer.} 
\newcommand\vEtype{Taut., Union, Piggy-back, Stored Proc., Infer., Alt. Encoding} 
\newcommand\vFtype{Union, Infer.,Alt. Encoding} 
\newcommand\vGtype{Infer.} 
\newcommand\vHtype{Union, Infer.} 
\newcommand\vItype{Infer.} 
\newcommand\vJtype{Second order inj.} 
\newcommand\vKtype{Infer., Alt. Encoding} 

\newcommand\vAdef{\vcG{}Yes}
\newcommand\vBdef{\vcG{}Yes}
\newcommand\vCdef{\vcG{}Yes}
\newcommand\vDdef{\vcG{}Yes}
\newcommand\vEdef{\vcG{}Yes}
\newcommand\vFdef{\vcG{}Yes}
\newcommand\vGdef{\vcG{}Yes}
\newcommand\vHdef{\vcG{}Yes}
\newcommand\vIdef{\vcG{}Yes}
\newcommand\vJdef{\vcG{}Yes}
\newcommand\vKdef{\vcG{}Yes}

\newcommand\vArules{{(update, wp\_em\_modals, none, [(=,field,literal),(IN,field,literal)])}}
\newcommand\vBrules{{(update, wp\_polls, none, [(=,field,lietral)])}}
\newcommand\vCrules{{(select, wp\_formmaker\_submits, and, [(=,field,literal)])}}
\newcommand\vDrules{{(select, wp\_posts, and, [(=,field,literal)])}}
\newcommand\vErules{{(select, users, and,[(=,field,literal)])}}
\newcommand\vFrules{{(select, [users, languages, fields], both, [(=,field,literal),(=, field, field),(IN, field, literal)])}}
\newcommand\vGrules{{(select, js\_jobs\_fieldsordering, none, [(=, field, literal)])}}
\newcommand\vHrules{{(select, jephotogallery, none, [(=, field, literal)])}}
\newcommand\vIrules{{(select, jquickcontanct\_captach, none, [(=, field, literal)])}}
\newcommand\vJrules{{(select, template\_styles, and, [(=, field, literal)])}}
\newcommand\vKrules{{(select, catalog\_product\_frontend\_action, and, [(>=, field, literal),(<=, field, literal)])}}
\begin{table*}[!h]
	\resizebox{\textwidth}{!}{
\begin{tabular}{|l|l|l|l|l|l|}
	\hline
	Application &  Vulnerability & SQLi Type & Available query descriptors\\ \hline
	\vAapp{} & \vAvuln{} & \vAtype{} & \vArules{}\\
	\vAapp{} & \vBvuln{} & \vBtype{} & \vBrules{}\\
	\vAapp{} & \vCvuln{} & \vCtype{} & \vCrules{}\\
	\vAapp{} & \vDvuln{} & \vDtype{} & \vDrules{}\\
	\vBapp{} & \vEvuln{} & \vEtype{} & \vErules{}\\
	\vCapp{} & \vFvuln{} & \vFtype{} & \vFrules{}\\
	\vDapp{} & \vGvuln{} & \vGtype{} & \vGrules{}\\
	\vDapp{} & \vHvuln{} & \vHtype{} & \vHrules{}\\
	\vDapp{} & \vIvuln{} & \vItype{} & \vIrules{}\\
	\vDapp{} & \vJvuln{} & \vJtype{} & \vJrules{}\\
	\vEapp{} & \vKvuln{} & \vKtype{} & \vKrules{}\\
	\hline
\end{tabular}
}
	\caption{Exploits blocked by \vP{}}
	\label{tab:defense}
\end{table*}

In Table \ref{tab:defense}, we present the list of SQLi vulnerabilities that
\vP{} defends the \vwebapp{}s against. 
The second column in Table \ref{tab:defense}, represents the ID assigned to
each vulnerability.
We marked the SQLi vulnerabilities that reside in the core of \vwebapp{}s by
\circled{C}. 
The third column shows the type of attacks we performed to exploit the
respective vulnerability. 
\vP{} protects the \vwebapp{}s against all 11 SQLi exploits in our dataset,
while SEPTIC can only defend against four SQLi exploits that only reside in
Wordpress plugins.
To evaluate the potential of circumventing \vP{}, we also listed the available
query descriptors for the SQL queries that the vulnerable PHP function in each
\vwebapp{} or plugin can issue.
For instance, any potential exploit against the first vulnerability in
Table~\ref{tab:defense} is restricted to an \texttt{UPDATE} query exclusively
on table \texttt{wp\_em\_modals} without further logical operators.
Furthermore, the exploit can only use SQL functions \texttt{"="} and
\texttt{"IN"}.
\subsection{Performance (RQ3)} 
Performance/responsiveness is a crucial factor for \vwebapp{}s.
Therefore, we evaluate \vP{}'s performance overhead.
In \vP{}, the first three steps can be performed offline. 
Steps \vsone{} and \vsthree{} are automatic and do not rely on help from the
administrator. 
In step \vstwo{}, the administrator must perform unit tests or create benign
traffic in the \vwebapp{} to train \vP{}.
Step \vsfour{} is deployed as a MySQL plugin and a set of modified PHP database
extensions to sandbox databases against malicious SQL queries. 
The MySQL server loads \vP{}'s protection plugin upon launch. 
\vP{} loads the profile and waits for incoming SQL queries.   
We perform our experiments on a 4-core Intel Core i7-6700 with 4Gb of memory
2133Mhz DDR4 that runs Linux 4.9.0, with Nginx 1.13.0, PHP 7.1.20, and MySQL
5.7.
For the performance evaluation, we created a Docker~\cite{docker}
container that runs with a default configuration of PHP, Nginx, and MySQL containing
the Drupal 7.0 \vwebapp{}.
We measure the performance overhead of \vP{} using
ApacheBench~\cite{apachebench}, a tool for benchmarking HTTP web servers.
We simulated a real-world scenario by increasing the level of concurrency in
ApacheBench.
The level of concurrency shows the number of open requests at a time.
We measured the network response time of \texttt{index.html} in Drupal 7.0 that
issues 26 queries to MySQL.
For more precise results, we measured the response time for 10,000 requests at
multiple levels of concurrency. 
Table~\ref{tab:overhead} presents our results for the aforementioned scenario.
The first column in Table~\ref{tab:overhead} shows the level of concurrency for
each test. 
The next two columns in Table~\ref{tab:overhead} present the network response
time for Drupal with/without \vP{}. 
As shown in Table~\ref{tab:overhead} \vP{} incurs less than (2.5\%) overhead to
the network response time of the server.
Based on the strong protections afforded by \vP{}, we consider this overhead
acceptable.
Furthermore, \vP{} is a prototype with no emphasis on performance optimization.
Such optimizations likely could reduce the overhead even further.
We also measured the execution time of queries in MySQL. 
We modified the source code of MySQL to calculate the time it takes for MySQL
to execute a SQL query. 
For this experiment, we used ApacheBench to send 10,000 requests to
\texttt{index.html} in Drupal 7.0, which issued a total of 260,000 queries to
MySQL.
We measured the average execution time of issued queries for  two different
scenarios. 
The first scenario is MySQL without \vP{}'s plugin, and in the second scenario,
we enabled \vP{}'s plugin in MySQL.
The last two columns in Table~\ref{tab:overhead} present the average execution
time of all the received queries to MySQL. 
The performance overhead of \vP{} in MySQL is less than 0.31 ms for each query. 
\begin{table}[H]
\resizebox{\columnwidth}{!}{
\begin{tabular}{crrrr}

	& \multicolumn{2}{c}{Server Response Time(ms)} & \multicolumn{2}{c}{MySQL Execution Time(ms)} \\\hline
	Concurrency  & Unprotected & Protected & Unprotected & Protected\\ \hline
	1 		& 27.792 	& 28.338 (1.96\%) & 0.150& 0.23\\\hline
	4 		& 11.644	& 11.813 (1.45\%) & 0.669&0.90\\ \hline 
	8 		& 8.907		& 9.127 (2.46\%) & 0.732&1.02\\ \hline
	16 		& 8.885 	& 9.084 (2.23\%) & 0.740&1.05\\ \hline
	32 		& 8.971 	& 9.182 (2.35\%) & 0.747&1.02\\\hline
\end{tabular}
}
	\caption{Response times for requests to Drupal index.php}
\label{tab:overhead}
\end{table}

\subsubsection{False Positive Evaluation}\label{subsec:fp}
We count an operation as a false positive if \vP{} blocks a benign query to the
database.
For the false positive evaluation, we evaluated \vP{} with Wordpress 4.7 and
Drupal 7.0.
For each \vwebapp{} we used the profile that we built in \texttt{RQ2}.
Then, we configured \vP{} in enforcement mode and replayed browsing traces
collected by Selenium~\cite{selenium}.
Our browsing traces explored the \vwebapp{} as a user and administrator with
the goal of covering the \vwebapp{} as much as possible.
Based on Table~\ref{tab:intersec}, only 10.11\% of the issued queries during
benign browsing and the unit test had the same query structure. 
This legitimate difference in the query structure of issued queries renders
prior approaches that build their profile based on query structure unable to
distinguish benign SQL queries from malicious ones.
For instance, SEPTIC has above 89\% false positive on the same test for Drupal
7.0.
\vP{} allows a query to execute in MySQL as long as the query matches at least
one of the query descriptors associated with the PHP function in the profile.
In the false positive test for Drupal, \vP{} did not block any query from the
benign Selenium browsing.
This shows that although the PHP functions during training and testing used
different queries, the query descriptors were the same.
Table~\ref{tab:intersec} shows that 82.57\% of queries in our benign browsing
were similar to queries recorded for \vP{}'s profile.
Although the rate of similar issued queries during training and testing of
Wordpress is higher that Drupal, \vP{} blocked 7 unique queries during the
benign browsing, which corresponds to 5\% of all issued queries.
%
%
There are two main reasons for the false positives in Wordpress.
The first reason is MySQL modifying the query based on the arguments passed to
SQL function in the query.
For instance, if the length of the array passed to the \texttt{IN} statement in
a query is one, MySQL modifies the \texttt{IN} statement to an \texttt{equal
(=)} statement.
This modification in the query and subsequently in the parse tree of the query
leads to false positives for \vP{} since \vP{} encounters a different function
in enforcement than what is in the profile.
The second reason is missing PHP functions in the profile.
During the enforcement, \vP{} blocks the SQL query if \vP{} does not find any
query descriptor for a PHP function that issued the SQL query.
Six out of seven false positives in Wordpress was due to lack of query
descriptors for the PHP function during benign browsing, which implies that
covering all the functions that can issue a query during the training is an
important factor for \vP{} (Discussed further in \S~\ref{sec:limit}).
%
\begin{table}[!h]
\resizebox{\columnwidth}{!}{
\begin{tabular}{ccccc}
	\vwebapp{} & Unit tests & Selenium  & (Unit tests $\cap$ Selenium) & False Positive\\\hline
	Drupal & 299961 & 336 & 34 (10.11\%) & 0\\
	Wordpress & 3099 & 132 & 109 (82.57\%) & 7\\ \hline
\end{tabular}
	}
	\caption{Number of unique SQL queries during unit testing and Selenium browsing }
\label{tab:intersec}

\end{table}

\subsection{Artifact Availability}

\vP{} implementation is open-source and available at
\url{https://www.github.com/BUseclab/SQLBlock}.
Additionally, we provide the five Docker containers that include a total of
\vdataset{} vulnerable PHP \vwebapp{}s and plugins that we used in our
evaluation. 
Our vulnerability dataset and the automated scripts were a significant part of
our evaluation, and we think that it can be useful for future works in this
area. 

\vspace*{-2mm}
\section{Discussion and Limitations}\label{sec:limit}
In this section, we discuss the limitations of the \vP{} and possible future
works in this area.
\textbf{eval Function: } 
PHP \vwebapp{}s use dynamic features implemented in PHP extensively, such as
the \texttt{eval} function, which evaluates a string argument as a PHP code.  
Currently, \vP{} does not handle function and class definitions inside
\texttt{eval}.
A \vwebapp{} can use \texttt{eval} for defining the \vdbapi{} or procedures
dynamically and use it across the \vwebapp{}. 
This leads to generating a non-complete list of PHP \vdbapi{} and interfaces
for a PHP \vwebapp{} in the step \vsone{}.
In such cases, \vP{} maps the query descriptors to a small set of PHP functions
that can allow the attacker to execute a malicious query.
In future work, the static analyzer in \vP{} can be improved to handle the
static PHP code passed to \texttt{eval}, to determine a more precise \vdbal{}.
\textbf{Incomplete coverage during training: }
PHP \vwebapp{}s generate dynamic queries based on user inputs. 
This approach makes it impossible to issue all possible queries to the database
during the training phase. 
Dynamic analyses suffer from incomplete training phases, and \vP{} is not an
exception.
Our Wordpress false positive test shows that the incomplete coverage of the
issued queries leads to \vP{} blocking benign queries.

\vspace*{-2mm}
\section{Conclusion}
We present \vP{}, a hybrid dynamic-static technique, to restrict the PHP
\vwebapp{}'s access to the database.
During the training step, \vP{} infers issued SQL queries and their respective
PHP \vcs{}s.
Using a lightweight static analysis, \vP{} extracts a list of \vdbapi{} and
procedures in the PHP \vwebapp{}. 
In the third step, \vP{} creates a set of query descriptors for each PHP
function in the PHP \vwebapp{} that issued a SQL query to the database. 
In the final step, \vP{} acts as a MySQL plugin to restrict the interaction of
the PHP \vwebapp{} and MySQL based on the generated query descriptors. 
\vP{} can prevent SQLi attacks against \vdataset{} vulnerabilities in the top
four most popular PHP \vwebapp{}s and seven plugins without any false positives
for Drupal 7.0 and a low number of seven false positives for Wordpress benign
browsing.
\section*{Acknowledgements}
We thank our shepherd Giancarlo Pellegrino and the anonymous reviewers for
their insightful comments and feedback.
This work was supported by the Office of Naval Research (ONR) under grant
N00014-17-1-2541.  \vspace*{-2mm} \balance

\bibliographystyle{ACM-Reference-Format}
\bibliography{paper}
\end{document}